\documentclass{article}
\usepackage[T1]{fontenc}
\usepackage[utf8]{inputenc}
\usepackage{ismir} 
\usepackage{amsmath,cite,url}
\usepackage{graphicx}
\usepackage{color}
\usepackage{booktabs}
\usepackage{pifont}
\usepackage{multirow}
\usepackage{amsmath}
\usepackage{amssymb}
\usepackage{makecell}

\title{Phase-Based Onset Enhancement for Piano Transcription in Ensemble Settings}

\multauthor
  {Hyemi Kim$^{1,2}$ \hspace{1cm} Jiyun Park$^3$ \hspace{1cm} Juhan Nam$^{2,3}$}
  {
  $^1$ Electronics and Telecommunications Research Institute (ETRI), Daejeon, South Korea\\    
  $^2$ Kim Jaechul AI Graduate School, KAIST, South Korea\\
  $^3$ Graduate School of Culture Technology, KAIST, South Korea\\
  {\tt\small miya0404@etri.re.kr, june@kaist.ac.kr, juhan.nam@kaist.ac.kr}
  }

\def\authorname{H. Kim, J. Park, and J. Nam}

\begin{document}

\maketitle

\begin{abstract}
Piano transcription models have achieved very high performance in recent state-of-the-art systems, but their performance tends to degrade in ensemble settings where the piano is played alongside other instruments.
While some approaches attempt to address this issue by incorporating multi-instrumental datasets, the problem remains challenging. 
In particular, when other instruments produce high-energy sustained tones while relatively low-energy piano notes occur simultaneously, the magnitude components of their frequency representations overlap, making it difficult to accurately distinguish piano notes.
To address this issue, we propose an improved piano transcription model that jointly leverages magnitude and phase information, enabling more robust piano transcription performance in ensemble settings. 
We evaluate the proposed model on two piano ensemble datasets with precise MIDI note annotations, which we obtained by revisiting and extending existing datasets.
Notably, we constructed a new piano concerto dataset, PCD-K, by playing a piano part on a Disklavier and recording the audio and MIDI pairs with pre-recorded orchestral tracks.  
Experimental results demonstrate that phase information is highly effective in improving performance, outperforming magnitude-based approaches in ensemble settings.
\end{abstract}

\begin{figure}
  \centering
  \includegraphics[width=\linewidth]{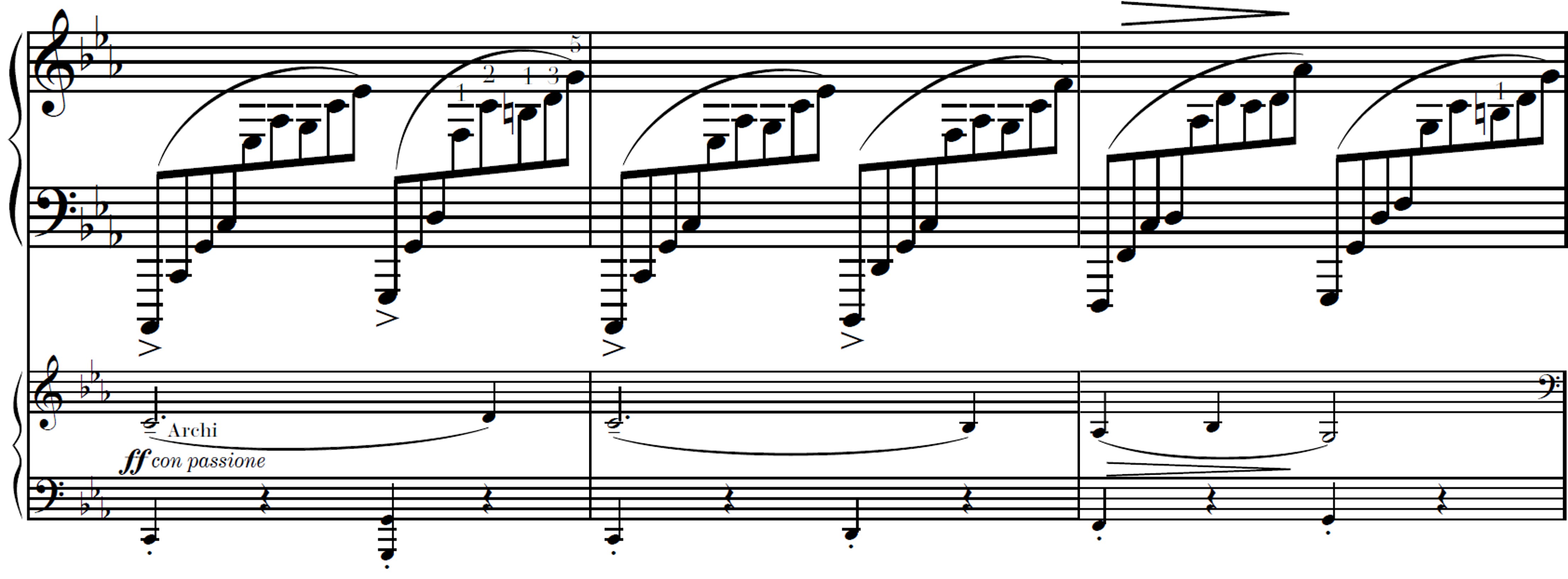}
  \vspace{-15pt}
  \caption{An excerpt from the score of Rachmaninoff’s Piano Concerto No.2, with piano (upper) and orchestra (lower).}
  \label{fig:score}
\end{figure}

\section{Introduction}\label{sec:introduction}
Piano transcription has achieved high performance through continuous advancements in neural network architectures \cite{transkunv1, transkun, onsetsandframes, kong2021high, jongwook, kwon, hppnet, wang_taslp, seq-to-seq, toyama2023hft, roll_lm}, as well as the availability of the MAESTRO dataset\cite{maestro}, a large-scale dataset with a high-precision MIDI capture system.
These studies are conducted on solo piano recordings, transcribing note onset, offset, pitch, and velocity, with some approaches further modeling expressive elements such as the sustain pedal \cite{kong2021high, transkunv1, transkun, wang_taslp}.
However, when the piano is played with other instruments in ensemble settings such as accompaniment, piano trios, or piano concertos with a full orchestra, transcription performance degrades significantly.

These limitations can be addressed by adopting a multi-instrument music transcription approach \cite{wu, mt3, multitrack_trans, YourMT3+}, which leverages datasets of multiple instruments and enables transcription of each instrument in ensemble settings.
Incorporating multi-instrument recordings into the training of piano transcription models can further improve their performance in ensemble scenarios.

In terms of input representations, existing piano transcription studies commonly use log-mel spectrograms or Constant-Q transforms\cite{cqt}. 
These representations rely primarily on magnitude information, which works effectively in solo piano settings.
However, in ensemble scenarios where the piano is played alongside louder instruments, its sound can be masked, and its onset signals become less distinguishable.
For example, in Figure \ref{fig:score}, when the orchestra performs loud, sustained tones at fortissimo while relatively low-energy piano onsets occur simultaneously or during these sustained passages, the characteristics of the piano onsets may not be sufficiently captured in magnitude-based representations.
This issue becomes more severe when the instruments share the same pitch or have overlapping harmonics.

Instrument onsets are typically associated not only with changes in magnitude\cite{sync_highres} but also with variations in phase across frequency components\cite{onsetdetection_bello, dixon2006onset}.
Phase information can be leveraged to more robustly detect piano onsets, even in ensemble settings where other instruments are significantly louder.
The orchestra maintains a loud sustained tone, as shown in Figure \ref{fig:score}, during which the phase in these frequency bands remains relatively stable.
However, such piano onsets induce abrupt phase variations within the same frequency bands.
While magnitude-based representations may fail to detect piano onsets when they are relatively weak, these phase variations can still provide a reliable cue.

Motivated by this observation, we propose an improved model that incorporates phase variations alongside magnitude-based features.
As the transcription backbone, we adopt the Transkun model~\cite{transkun}, a semi-CRF-based piano transcription framework that achieves state-of-the-art performance while also supporting pedal transcription, which is not typically handled in multi-instrument transcription approaches.
The Symphony dataset is constructed and incorporated as data augmentation in line with multi-instrument transcription approaches.

Evaluating the proposed model in such ensemble settings requires appropriate datasets.
Conventional piano transcription datasets~\cite{maestro, maps, smd}, which focus on solo piano, are not suitable for this purpose.
Upon revisiting the MAESTRO dataset, commonly regarded as a solo piano dataset, we find that it also contains ensemble recordings with strings, which we use as our first evaluation dataset.
In addition, we collected a new piano concerto dataset with precise piano MIDI annotations, PCD-K, as another test set. This dataset is a variation of the Piano Concerto Dataset (PCD)~\cite{pcd} that provides separate audio recordings for the piano and the full orchestra along with corresponding scores and note annotations. Since the note annotations are obtained via audio–MIDI alignment, they lack sufficient accuracy for evaluating transcription models. To address the issue, we invited pianists and recorded their performances on a Disklavier piano that allows for capturing both audio and MIDI.

In summary, the main contributions of this paper are as follows.
We propose a novel method that incorporates phase information, a relatively unexplored aspect in piano transcription, to improve the performance in ensemble settings, built upon a state-of-the-art transcription model. 
We further identify and construct evaluation datasets for classical piano trios and concertos, respectively, and compare note and sustain pedal transcription performance against conventional approaches.

\begin{figure*}
  \centering
  \includegraphics[alt={Overview of the proposed model},width=0.9\linewidth]{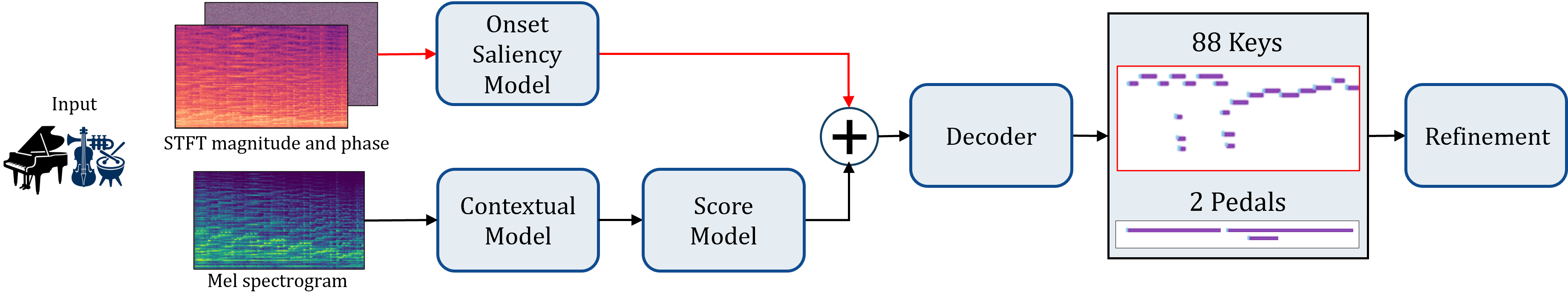}
  \vspace{-5pt}
  \caption{Overview of the proposed model. The STFT magnitude and phase are used to enhance interval scoring for the 88 piano keys, as indicated by the red arrows and the red boxed region.} 
  \label{fig:system}
\end{figure*}

\section{Related works}

Early neural approaches to piano transcription predominantly operate at the frame level, estimating note-related signals such as onset, offset, and frame-wise pitch activations, followed by post-processing steps to reconstruct note events. Onsets and Frames~\cite{onsetsandframes} introduced a model that explicitly separates onset and frame predictions, and subsequent works have extended this framework by incorporating onset and offset regression~\cite {kong2021high}, adversarial learning~\cite {jongwook}, and note-state modeling and autoregressive connections~\cite {kwon}. 
CQT-based input representations have also been explored in conjunction with harmonic priors~\cite{hppnet} and multitask learning with frequency–time feature modulation~\cite{wang_taslp}.

Beyond frame-based approaches, alternative formulations have been explored to more directly model note structures and temporal dependencies. 
These include transcription as translation from audio to symbolic token sequences~\cite{seq-to-seq}, hierarchical time–frequency attention~\cite{toyama2023hft}, and language model-based note generation conditioned on piano roll representations~\cite{roll_lm}.
Structured approaches~\cite{transkunv1, transkun} also model notes explicitly as time intervals.

The methods discussed above rely on magnitude-based time–frequency representations, such as mel-spectrograms or constant-Q transforms (CQT), as input features, and do not explicitly leverage phase information.

Available datasets also present important limitations.
Piano transcription datasets \cite{maestro, maps, smd} provide highly accurate labels through a capture system, but are limited to solo performances.
In contrast, ensemble datasets that include piano cover piano ensembles \cite{musicnet, trios}, piano concertos \cite{pcd}, and broader classical music collections \cite{rwc}, 
but rely on score-aligned note annotations, which leads to lower accuracy.
These limitations point to the need for a piano ensemble dataset with precise transcription labels.

\section{Proposed method}

\subsection{Revisiting Transkun}
The Transkun model~\cite{transkunv1, transkun}, commonly referred to as the semi-CRF model, formulates piano transcription as an interval-based structured prediction problem using a neural semi-Markov conditional random field.
Rather than estimating onset, offset, and frame-level activations separately, the model assigns scores to all possible time intervals, treating each interval as a candidate musical event.
The resulting scores define a probabilistic model over sequences of non-overlapping events, and the model is trained by maximizing the conditional likelihood with additional attribute losses for velocity estimation and refinement of onset and offset positions.
This formulation enables direct optimization of note-level events without post-processing. 
In \cite{transkun}, the MLP-based interval scoring is replaced with a scaled inner product formulation to improve computational efficiency, which aligns with attention mechanisms. 
The model further adopts an encoder-only transformer backbone accordingly.

\subsection{Transkun-Phase}

We incorporate phase-based information to enhance onset saliency in the interval score map, improving the detectability of piano onsets under interference from other instruments.
Figure \ref{fig:system} illustrates the overall system.
The STFT $X(f,t)$ of the input audio can be expressed in polar form as
$
X(f,t) = |X(f,t)| e^{j\phi(f,t)},
$
where $|X(f,t)|$ and $\phi(f,t)$ denote the magnitude and phase at frequency bin $f$ and time frame $t$, respectively.
The phase difference is then measured using the phase component of the complex STFT.
\begin{equation}\label{}
\Delta \phi(f,t)
=\arg\!\left(
e^{j\left(\phi(f,t)-\phi(f,t-1)\right)}
\right)
\end{equation}
Onset saliency is computed as
\begin{equation}\label{onset_saliency}
S(f,t)=|\Delta \phi(f,t)|\cdot |X(f,t)|^{\gamma}
\end{equation} 
where $S \in \mathbb{R}^{F \times T}$, with the magnitude controlled by a scaling factor $\gamma$ to suppress noise-induced fluctuations and emphasize true note onset events, rather than relying solely on phase changes.
We estimate an onset probability for each frame $i$ and pitch $p$ using an onset prediction head
\begin{equation}\label{}
p_{\mathrm{onset}}(i,p)=\sigma\left(\mathrm{OnsetHead}(S)_{i,p}\right)
\end{equation}
where $\mathrm{OnsetHead}(\cdot)$ denotes a multi-layer perceptron (MLP) consisting of a LayerNorm followed by two linear layers with a GELU activation in between, producing onset logits for each piano key.

Score maps are predicted for both the 88 piano keys and the two pedal controls in \cite{transkunv1, transkun}.
We apply our onset saliency enhancement only to the key score maps, excluding those of the pedal controls.
This is because note onsets introduce new excitations with arbitrary initial phases, leading to abrupt phase discontinuities, whereas pedal events only modify the decay characteristics of ongoing vibrations, preserving phase continuity.

The final interval score is defined as the sum of the interval score derived from the mel-spectrogram representation
and the phase-based onset score, 
\begin{equation}\label{}
score(i,j,p)=
score_{\mathrm{mel}}(i,j,p)
+
\lambda_{\mathrm{phase}}score_{\mathrm{phase}}(i,p)
\end{equation}
where $\lambda_{\mathrm{phase}}$ controls the contribution of the phase-based term. 
The phase-based onset score is defined as
\begin{equation}\label{equ:phase_onset_score}
score_{\mathrm{phase}}(i,p)
=\log\Big(
p_{\mathrm{onset}}(i,p)+\varepsilon
\Big)
\end{equation}
where $\varepsilon$ is a small constant for numerical stability.

Since $score_{\mathrm{phase}}(i,p)$ depends only on the onset frame $i$, it is shared across all intervals $(i,j)$ with the same start frame, increasing the scores of intervals aligned with likely onset positions.

The training objective combines the CRF loss and attribute losses from \cite{transkun} with the phase-based onset loss
\begin{equation}\label{}
\mathcal{L} = \mathcal{L}_{\text{CRF}} + \mathcal{L}_{\text{attributes}} + \mathcal{L}_{\text{onset}}
\end{equation}
where $\mathcal{L}_{\text{CRF}}$ maximizes the conditional likelihood of the ground-truth event sequence under the semi-CRF formulation, and $\mathcal{L}_{\text{attributes}}$ corresponds to auxiliary losses for predicting event-level attributes, including velocity, refined onset, and refined offset.
The additional onset loss $\mathcal{L}_{\text{onset}}$ is defined as a binary cross-entropy loss between the predicted onset probability $p_{\mathrm{onset}}(i,p)$ and the ground-truth onset label.

\begin{table}[h]
\setlength{\tabcolsep}{3pt}
\resizebox{\columnwidth}{!}{
\begin{tabular}{lllc}
\toprule
Title & Composer & Split & Dur. \\
\midrule
Piano Trio in D Minor, Op.\ 32 & Anton Arensky & Train & 28:57 \\
Piano Trio in D Minor, Op.\ 32 & Anton Arensky & Train & 26:54 \\
Piano Trio in D Minor, Op.\ 32 & Anton Arensky & Train & 28:35 \\
Piano Trio in D Major, Op.\ 70 No.\ 1 & Ludwig van Beethoven & Train & 20:55 \\
Piano Trio in D Major, Op.\ 70 No.\ 1 & Ludwig van Beethoven & Train & 19:41 \\
Piano Trio No.\ 1 in D Major, Op.\ 49 & Felix Mendelssohn & Test & 28:42 \\
\bottomrule
\end{tabular}
}
\caption{The list of Piano Trios in the MAESTRO v2 dataset. (durations in mm:ss)}
\label{tab:maestrov2trio_list}
\end{table}

\begin{figure*}
  \centering
  \includegraphics[alt={ISMIR 2025 template example image},width=0.98\linewidth]{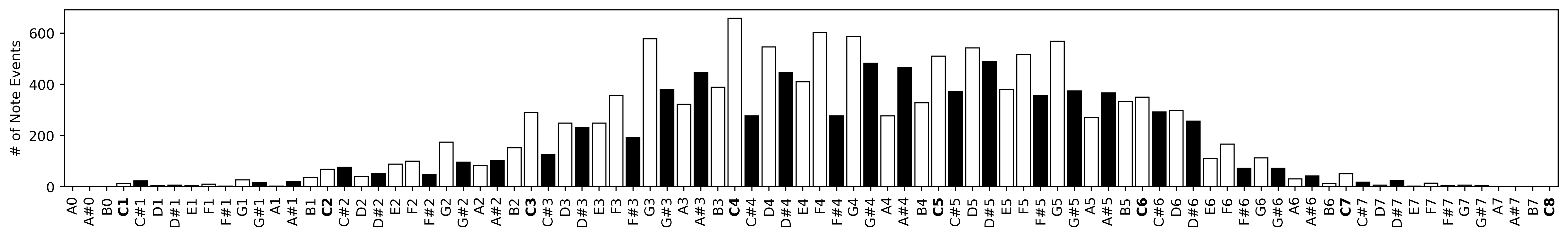}
  \vspace{-7pt}
  \caption{Pitch histogram of piano notes in the PCD-K dataset}
  \label{fig:pitch_histogram}
\end{figure*}

\section{Evaluation Datasets}

\subsection{MAESTRO v2 Trio}
We revisit the MAESTRO v2 dataset, primarily used for solo piano transcription, which includes six piano trio recordings across the training and test splits.
These recordings consist of piano, violin, and cello, include piano note and pedal annotations, and were unintentionally introduced in the process of expanding MAESTRO v1.
Those recordings were later removed, and the dataset was re-released as MAESTRO v3, which contains only solo piano recordings.
In this work, we regard these six recordings as a valuable test set for evaluating piano transcription in ensemble settings.
The titles, composers, and durations of the recordings are listed in Table \ref{tab:maestrov2trio_list}.
The total duration of this subset is approximately 2 hours and 33 minutes.

\subsection{PCD-K: Piano Concerto Dataset for Piano Transcription}

The Piano Concerto Dataset (PCD)~\cite{pcd} consists of 81 carefully selected 12-second multitrack excerpts from 15 piano concertos. 
It provides separate piano and orchestral audio stems, their mixtures, and corresponding scores and synchronized MIDI files.
Piano recordings were captured across four distinct acoustic environments, while orchestra recordings were sourced from the Music Minus One (MMO) collection\footnote{\url{https://www.halleonard.com/series/MMONE}}, which provides orchestral accompaniment tracks for piano concertos.
This is publicly available dataset\footnote{\url{https://audiolabs-erlangen.de/resources/MIR/PCD}}
that provides piano note annotations in the context of piano concertos.
However, since the MIDI was obtained via alignment \cite{muller2021sync} rather than captured with a high-precision sensor system that detects key movements, it is insufficiently accurate for transcription tasks.

To address this, we collected the piano part again by re-performing the pieces on a Disklavier piano and recording both audio and performance MIDI data simultaneously.
We conducted the recording in a soundproof room, using two microphones placed above the piano with the lid open. 
All piano performances were performed by eight researchers in the department that the authors are affiliated with. Two of them majored in piano performance. 
Performers listened to the 12-second orchestral track from the PCD dataset with headphones and performed in synchronization. 
This setup prevented audio bleed and allowed for clean isolation of the piano recording. 
A count-in beep, added in Logic Pro when necessary, was used to cue the performer at the start of the orchestra recording.
If any misalignment in synchronization was observed, manual adjustments were made to ensure accurate temporal alignment.
Each piano–orchestra mixture was created by balancing the two signals at equal levels.

Among the 81 excerpts in the PCD~\cite{pcd}, some include up to two alternate versions recorded in different locations.
In contrast, our dataset was recorded in a single environment, with each excerpt having only one version, yielding a total of 59 excerpts.
We also recorded 11 new excerpts from Rachmaninoff’s Rhapsody on a Theme of Paganini, Op. 43 (Var.18) and Chopin’s Piano Concerto No.1 in E minor, Op. 11 (Mov.1), which are not included in the PCD. 
Following the dataset construction pipeline of the PCD, we built a dataset comprising a total of 70 excerpts.
To reflect acoustic variability, the PCD includes both non-reverb and reverb versions of the same performance. 
In line with this design, the sound engineer responsible for the PCD produced a reverberant version of our recordings.
The piano note pitch histogram is shown in the Figure \ref{fig:pitch_histogram}.
The dataset is publicly available on Zenodo.\footnote{\url{https://zenodo.org/records/21528883}}

\section{Experimental Settings}

\subsection{Symphony: Additional Training Dataset}
To improve transcription performance in ensemble settings such as piano trios and concertos, 
the piano track from MAESTRO v3 is randomly mixed with an orchestral track during training, with random gain applied to control relative levels.
We construct the Symphony dataset that excludes the piano and covers a wide range of orchestral instruments, including strings, woodwinds, brass, and percussion.
Symphony recordings were collected from YouTube and IMSLP\footnote{\url{https://imslp.org/}}, covering 20 pieces by 12 composers and supplemented by a substantial collection of Mozart symphonies, for a total of 13 composers.
Total duration is approximately 25 hours.
The training and validation sets are split in a 9:1 ratio.

\subsection{Model Specification}

We compute the STFT of the mixture at a sampling rate of 44.1 kHz with a window size of 4096 and a hop size of 1024, and use both magnitude and phase components. 
Log mel-spectrograms with 229 mel bins spanning 30–8000 Hz are derived and used as input representations commonly adopted in existing methods.
Each audio sample is processed in 16-second segments.

We follow the training setup of the Transkun model, including the optimizer, learning rate schedule, and gradient clipping strategy. Specifically, we use the AdaBelief optimizer~\cite{adabelief} with a maximum learning rate of $4 \times 10^{-4}$ and a weight decay of $1 \times 10^{-2}$. The learning rate follows a cosine annealing schedule over the total number of iterations, with a 5\% warm-up phase at the beginning of training.
Gradient clipping is used to stabilize training, with the clipping threshold set to the 80\% quantile of gradient norms over the past 10,000 iterations.

The phase-based onset weight $\lambda_{\text{phase}}$ is set to 100 based on validation experiments to match the scale of the interval scores, and $\gamma$ in \eqnref{onset_saliency} is set to 1.
For validation, we adopt a deterministic procedure, generating the mixtures once and keeping them fixed across iterations.
The checkpoint with the highest note onset F1-score on the validation set is selected for evaluation.
Training is conducted for 500K iterations with a batch size of 8, taking approximately 6 days on a single NVIDIA RTX 4090.

\begin{table}[t]
\centering
\resizebox{\columnwidth}{!}{
\begin{tabular}{l|l|l}
\toprule
\textbf{Model} & \textbf{Additional Datasets} & \textbf{Input} \\
\midrule
Lowerbound Transkun\cite{transkun} &  $-$ & Mixture \\
Kong\cite{kong2021high} & Piano Trio & Mixture \\
YourMT3+\cite{YourMT3+} & Multi Inst. & Mixture \\
Transkun\_Aug\cite{transkun} & Environmental Sound & Mixture \\
\midrule
Transkun\cite{transkun} & Symphony & Mixture \\
Transkun-Phase (Ours) & Symphony & Mixture \\
\midrule
Upperbound Transkun\cite{transkun} & $-$ & Piano Solo \\
\bottomrule
\end{tabular}
}
\caption{Comparison of models with their additional training datasets and input audio type at inference time.}
\label{tab:models_aug_dataset}
\end{table}

\begin{table*}[t]
\centering
\setlength{\tabcolsep}{5pt}
\resizebox{\textwidth}{!}{
\begin{tabular}{l|ccc|ccc|ccc|ccc|ccc|ccc}
\toprule
\textbf{} 
& \multicolumn{12}{c|}{\textbf{Piano Note Transcription}} 
& \multicolumn{6}{c}{\textbf{Sustain Pedal Transcription}} \\
\cmidrule(lr){2-13} \cmidrule(lr){14-19}
\textbf{Method} 
& \multicolumn{3}{c|}{Frame} 
& \multicolumn{3}{c|}{Onset} 
& \multicolumn{3}{c|}{On+Off} 
& \multicolumn{3}{c|}{On+Off+Vel}
& \multicolumn{3}{c|}{Frame}
& \multicolumn{3}{c}{Onset} \\
\cmidrule(lr){2-4} \cmidrule(lr){5-7} \cmidrule(lr){8-10} \cmidrule(lr){11-13}
\cmidrule(lr){14-16} \cmidrule(lr){17-19}

& P & R & F1 & P & R & F1 & P & R & F1 & P & R & F1
& P & R & F1 & P & R & F1 \\
\midrule
Lowerbound & 74.45 & 82.51 & 78.13 & 56.67 & 90.98 & 69.81 & 43.42 & 69.60 & 53.46 & 37.18 & 59.54 & 45.76 & 92.53 & 67.51 & 77.83 & 43.57 & 44.15 & 43.79 \\
YourMT3+\cite{YourMT3+} & 82.72 & 78.88 & 80.71 & 91.06 & 85.38 & 88.12 & 68.48 & 64.24 & 66.28 & 30.54 & 28.66 & 29.56 & - & - & - & - & - & - \\
Transkun\_Aug\cite{transkun} & 85.49 & 86.62 & 86.01 & 88.83 & 90.61 & 89.69 & 75.52 & 77.11 & 76.29 & 73.19 & 74.72 & 73.92 & 94.03 & 87.99 & 90.88 & 75.72 & 62.99 & 68.53 \\
\midrule
Transkun\cite{transkun} & 87.66 & 91.63 & 89.54 & 97.73 & 89.84 & 93.61 & 85.82 & 78.96 & 82.24 & 84.36 & 77.62 & 80.84 & 94.97 & 89.89 & 92.34 & 75.13 & 60.47 & 66.65 \\
\textbf{Transkun-Phase (Ours)} & 89.10 & 92.66 & \textbf{90.82} & 98.02 & 91.60 & \textbf{94.69} & 86.84 & 81.22 & \textbf{83.93} & 85.64 & 80.11 & \textbf{82.78} & 94.24 & 91.82 & \textbf{93.00} & 76.04 & 63.55 & \textbf{68.97} \\

\bottomrule
\end{tabular}
}
\vspace{-5pt}
\caption{Piano note and sustain pedal transcription results on the MAESTRO v2 Trio subset}
\label{tab:results_MAESTROv2Trio_NotePedal}
\end{table*}


\begin{table*}[t]
\centering
\setlength{\tabcolsep}{5pt}
\resizebox{\textwidth}{!}{
\begin{tabular}{l|ccc|ccc|ccc|ccc|ccc|ccc}
\toprule
\textbf{} 
& \multicolumn{12}{c|}{\textbf{Piano Note Transcription}} 
& \multicolumn{6}{c}{\textbf{Sustain Pedal Transcription}} \\
\cmidrule(lr){2-13} \cmidrule(lr){14-19}

\textbf{Method} 
& \multicolumn{3}{c|}{Frame} 
& \multicolumn{3}{c|}{Onset} 
& \multicolumn{3}{c|}{On+Off} 
& \multicolumn{3}{c|}{On+Off+Vel}
& \multicolumn{3}{c|}{Frame}
& \multicolumn{3}{c}{Onset} \\
\cmidrule(lr){2-4} \cmidrule(lr){5-7} \cmidrule(lr){8-10} \cmidrule(lr){11-13}
\cmidrule(lr){14-16} \cmidrule(lr){17-19}

& P & R & F1 & P & R & F1 & P & R & F1 & P & R & F1
& P & R & F1 & P & R & F1 \\
\midrule
Lowerbound& 59.60 & 76.52 & 65.93 & 50.63 & 93.09 & 64.19 & 35.18 & 62.66 & 44.18 & 26.78 & 46.84 & 33.45 & 87.85 & 80.90 & 83.31 & 49.55 & 57.03 & 51.33 \\
Kong & 48.01 & 82.03 & 58.81 & 67.05 & 87.67 & 74.89 & 44.25 & 57.01 & 49.18 & 33.67 & 43.23 & 37.40 & 81.65 & 85.15 & 82.01 & 47.17 & 60.32 & 51.57 \\
YourMT3+\cite{YourMT3+} & 73.89 & 62.39 & 66.24 & 85.42 & 85.26 & 85.04 & 45.08 & 44.81 & 44.81 & 21.79 & 21.60 & 21.63 & - & - & - & - & - & - \\
Transkun\_Aug\cite{transkun} & 76.61 & 77.34 & \textbf{76.02} & 79.60 & 92.10 & 84.76 & 62.36 & 71.97 & 66.37 & 50.91 & 58.25 & 53.98 & 87.82 & 87.32 & \textbf{86.33} & 61.97 & 65.28 & \textbf{62.12} \\
\midrule
Transkun\cite{transkun} & 84.89 & 66.31 & 73.04 & 96.82 & 81.60 & 87.93 & 74.52 & 63.53 & 68.14 & 57.97 & 49.09 & 52.82 & 88.61 & 74.71 & 78.95 & 61.35 & 47.50 & 51.83 \\
\textbf{Transkun-Phase (Ours)} & 83.90 & 71.18 & 75.68 & 95.38 & 86.87 & \textbf{90.46} & 74.12 & 68.58 & \textbf{70.97} & 58.01 & 53.51 & \textbf{55.48} & 88.77 & 82.27 & 84.44 & 61.85 & 55.66 & 57.83 \\
\midrule
Upperbound & 90.02 & 87.29 & 88.35 & 98.52 & 97.30 & 97.87 & 89.43 & 88.38 & 88.87 & 73.68 & 72.73 & 73.18 & 96.05 & 93.57 & 94.45 & 85.50 & 86.48 & 85.38 \\

\bottomrule
\end{tabular}
}
\vspace{-5pt}
\caption{Piano note and sustain pedal transcription results on the PCD-K dataset}
\label{tab:results_PCD-K_NotePdeal}
\end{table*}

\subsection{Model Comparison}

We compare three publicly available pretrained models for piano transcription: Transkun\cite{transkun} and Kong\cite{kong2021high}, both supporting note and pedal prediction, and YourMT3+\cite{YourMT3+}, which is designed for multi-instrument transcription.
All three models are based on the MAESTRO dataset, but use different MAESTRO versions or are augmented with other datasets.

The comparison configurations are summarized in Table \ref{tab:models_aug_dataset}.
Two pretrained versions of Transkun \cite{transkun} are used, provided by the official repository\footnote{\url{https://github.com/yujia-yan/transkun}}. One is trained solely on MAESTRO v3 and serves as reference performance bounds, providing both a lower bound when evaluated on audio mixtures and an upper bound when evaluated on isolated piano tracks. 
The other, referred to as Transkun\_Aug, is trained with additional data augmentation using the ESC dataset \cite{esc} to enhance piano transcription performance.

Kong's piano transcription model\cite{kong2021high}\footnote{\url{https://github.com/qiuqiangkong/piano\_transcription\_inference}}, trained on MAESTRO v2, 
has already seen five of the six pieces in the MAESTRO v2 Trio subset during training and is therefore excluded from its evaluation to ensure a fair comparison.
YourMT3+\cite{YourMT3+}\footnote{\url{https://huggingface.co/spaces/mimbres/YourMT3}}, designed for multi-instrument transcription, is trained on diverse datasets covering piano \cite{maestro}, orchestral and ensemble instruments \cite{urmp, musicnet}, synthetic multi-instrument audio \cite{slakh}, guitar\cite{guitarset}, bass\cite{smtbass}, drums\cite{enstdrums, egmd}, and singing voice \cite{mirst500}.

We compare our method with Transkun\cite{transkun} when both are trained under the same settings on the MAESTRO and Symphony datasets, as well as with the aforementioned pretrained models.
Standard frame-based and note-based metrics from \textit{mir\_eval} \cite{mir_eval} are used, with an onset tolerance of 50 ms, an offset tolerance of the maximum of 50 ms or 20\% of the note duration, and a velocity tolerance of 0.1, following common evaluation settings in piano transcription.
During both training and inference, note offsets are extended using sustain pedal information.

\section{Experimental Results}
\subsection{Evaluation on MAESTRO v2 Trio}
In Table \ref{tab:results_MAESTROv2Trio_NotePedal}, the lowerbound Transkun model exhibits low precision but high recall in the note-onset metric, suggesting that it reliably captures most piano note onsets, while also producing a large number of false positives, which may be attributed to difficulty in distinguishing piano from other concurrently sounding instruments.
Due to the use of additional training datasets, both YourMT3+\cite{YourMT3+} and Transkun\_Aug\cite{transkun} achieve higher precision, resulting in improved note onset F1-scores, with Transkun\_Aug consistently outperforming YourMT3+, which exhibits substantially poorer performance on the note onset with offset and velocity metric.

The Transkun model, used as our baseline and trained on both the MAESTRO and Symphony datasets, shows an improvement in precision, achieving F1-scores of 93.6\%, 82.2\%, and 80.8\% for note onset, note onset with offset, and note onset with offset and velocity, respectively.
The proposed method outperforms all other models, achieving F1-scores of 94.7\%, 83.9\%, and 82.8\%.
It also achieves the best performance in sustain pedal transcription, with F1-scores of 93.0\% at the frame level and 69.0\% for onset.

\subsection{Evaluation on PCD-K}
Table \ref{tab:results_PCD-K_NotePdeal} shows the results on PCD-K.
Similar to Table \ref{tab:results_MAESTROv2Trio_NotePedal}, the lower bound shows low precision of 50.6\%. While pretrained models trained with additional datasets show some improvement in precision. 
Among the pretrained models, Transkun\_Aug\cite{transkun} achieves the best overall performance, particularly on the note-onset-with-offset and frame-level metrics. For the note-onset metric, however, YourMT3+\cite{YourMT3+} slightly outperforms Transkun\_Aug. Kong follows Transkun\_Aug on the note-onset-with-offset metrics.

The Transkun model, trained on both the MAESTRO and symphony datasets, shows an improvement in precision, achieving F1-scores of 87.9\%, 68.1\%, and 52.8\% for note onset, note onset with offset, and note onset with offset and velocity, respectively.
The proposed method further improves the results, with higher scores of 90.5\%, 71.0\%, and 55.5\%, demonstrating a substantial gain.

In contrast, Transkun\_Aug achieves the best performance across both frame-level and sustain pedal prediction.
This can be attributed to the fact that sustain pedal signals depend on subtle energy variations such as decay and reverberation, for which data augmentation using the environmental sound dataset\cite{esc} improves robustness, especially under out-of-distribution conditions.

\begin{figure}
  \centering
  \includegraphics[width=\linewidth]{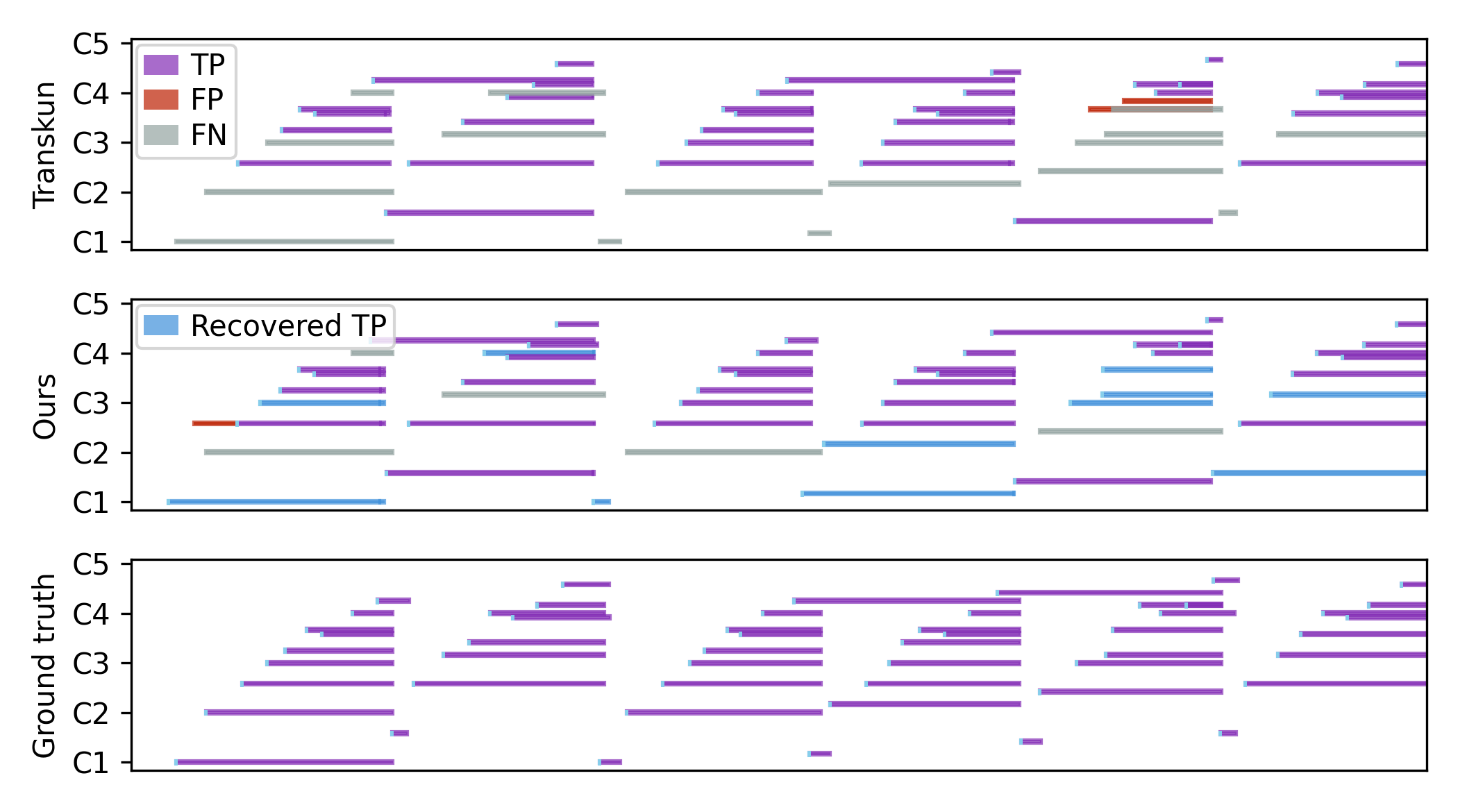}
  \vspace{-20pt}
  \caption{An example of piano transcription results (the Rachmaninoff’s Piano Concerto No.2 excerpt in Figure \ref{fig:score}). 
  True positives (TP), false positives (FP), and false negatives (FN) are shown in purple, red, and gray, respectively, while recovered true positives are highlighted in sky blue.}  
  \label{fig:midi_result}  
\end{figure}

Experimental results show that the proposed method achieves higher overall F1-scores than the Transkun model, with consistently significant improvements in recall across both datasets.
An example of piano transcription results for an excerpt from Rachmaninoff’s Piano Concerto No. 2 (Figure \ref{fig:score}) is shown in Figure~\ref{fig:midi_result}, highlighting these recall improvements.
Note intervals whose onsets are matched within the onset tolerance are counted as true positives (TP), and the proposed method recovers more true positives missed by the baseline, as indicated in sky blue.

\subsection{Ablation study}

We conduct an ablation study to investigate the effect of phase information on onset saliency estimation, comparing the combination of magnitude and first-order phase difference ($A \times |\Delta \phi|$), as defined in \eqnref{onset_saliency}, with additional variants, including the second-order phase difference ($A \times |\Delta^2 \phi|$) and the magnitude-only variant ($A$).
The second-order phase difference is computed as
\begin{equation}
\Delta^2 \phi(f,t)
=\arg\left(
e^{j\left(\phi(f,t+1)-2\phi(f,t)+\phi(f,t-1)\right)}
\right).
\end{equation}

As shown in Table~\ref{tab:results_mag_and_phase}, the combination of magnitude and first-order phase difference achieves the best performance on the MAESTRO v2 Trio subset, with the second-order variant showing comparable performance, while the magnitude-only variant exhibits a clear performance drop. 
On the PCD-K dataset, this trend becomes less clear, but the method incorporating the first-order phase difference still achieves the best overall performance.
The magnitude-only variant shows a moderate drop in note onset F1-score and a clear drop in pedal onset F1-score.
Overall, these results indicate that incorporating phase information improves transcription performance.

However, an exception is observed in the note onset F1-score with offset and velocity, where the magnitude-only model achieves a slightly higher score.
This discrepancy may stem not only from inherent limitations of velocity evaluation, but also from distribution shifts under out-of-distribution conditions.
Velocity estimation is inherently ill-defined in polyphonic settings, as note-level loudness depends on complex musical context, and MIDI velocity does not directly correspond to perceived loudness \cite{kim2024velocity}. Furthermore, velocity values may vary across recordings, introducing dataset-dependent variations that the evaluation metric does not fully capture.

\begin{table}[t]
\centering
\setlength{\tabcolsep}{4pt}
\resizebox{\columnwidth}{!}{
\begin{tabular}{l|l|cccc|cc}
\toprule
\textbf{}&\textbf{}
& \multicolumn{4}{c|}{\textbf{Piano Note}}
& \multicolumn{2}{c}{\textbf{Sustain Pedal}} \\
\cmidrule(lr){3-6} \cmidrule(lr){7-8}
\textbf{Dataset}&\textbf{Input}
& \textbf{Fr} 
& \textbf{O} 
& \textbf{OF} 
& \textbf{OFV}
& \textbf{Fr} 
& \textbf{O} \\
\midrule
\multirow{3}{*}{\makecell[l]{MAESTRO\\v2 Trio}}
&$A \times |\Delta \phi|$  
& \textbf{90.82} & \textbf{94.83} & \textbf{84.33} & \textbf{83.27} & \textbf{93.00} & \textbf{69.46} \\
&$A \times |\Delta^2 \phi|$  
& 90.50 & 94.58 & 84.13 & 83.13 & 92.54 & 68.59 \\
&$A$
& 90.18 & 93.96 & 83.03 & 81.96 & 91.39 & 66.94 \\
\midrule
\multirow{3}{*}{PCD-K}
&$A \times |\Delta \phi|$  
& \textbf{75.68} & \textbf{90.33} & \textbf{71.11} & 57.41 & \textbf{84.44} & \textbf{63.31} \\
&$A \times |\Delta^2 \phi|$  
&74.81 & 89.66 & 69.98 & 57.27 & 82.86 & 63.10 \\
& $A$
&75.04 & 88.04 & 70.42 & \textbf{57.85} & 82.85 & 58.95 \\

\bottomrule
\end{tabular}
}
\vspace{-5pt}
\caption{Ablation results on STFT input representations measured by F1-score. Fr, O, OF, and OFV denote frame, onset, onset with offset, and onset with offset and velocity, respectively.}
\label{tab:results_mag_and_phase}
\end{table}

\vspace{-7pt}
\section{Conclusions}
In this paper, we propose a novel method that incorporates phase information, a relatively underexplored aspect in piano transcription, and show that it improves performance in ensemble settings, particularly in the classical concert repertoire such as piano trios and concertos.
We also prepared two evaluation datasets. 
One is obtained by identifying a subset of piano trio recordings within the MAESTRO v2 dataset for piano transcription in ensemble settings.
The other is constructed by re-performing and re-recording the piano tracks of the PCD dataset using a piano equipped with MIDI capture, thereby creating the PCD-K dataset with accurate piano labels.
Experimental results indicate that the proposed method outperforms the baseline model trained on the same additional Symphony dataset, achieving higher F1-scores for both note and pedal transcription, with notable gains in recall.
We demonstrate that the proposed method improves transcription performance by enhancing the onset saliency of piano notes masked by high-energy sustained sounds from other instruments.
The proposed approach may extend to broader multi-instrument music transcription and score-following tasks.
Our dataset can also serve as a new benchmark for multi-instrument transcription by providing accurate piano annotations in multi-instrument settings 
and for piano source separation.

\section{AI Usage Statement}
A large language model (LLM) was used to assist in categorizing prior work in the literature review based on papers selected by the authors.

\section{Acknowledgments}
We thank Dr. Simon Schw\"ar for professionally processing the reverberation in our dataset, consistent with his previous work on the PCD.

This work was supported by the Culture, Sports and Tourism R\&D Program through the Korea Creative Content Agency grant funded by the Ministry of Culture, Sports and Tourism (Development of Core Technologies for Copyright Verification of AI-Generated and Deepfake Music, RS-2025-02216483); by the National Research Foundation of Korea (NRF) grant funded by the Korea government (MSIT) (RS-2023-NR077289); and by Institute for Information \& Communications Technology Planning \& Evaluation (IITP) grant funded by the Korea government (MSIT) (RS-2019-II190075, Artificial Intelligence Graduate School Program (KAIST)) (Contribution rate: 33.3\% each).

\bibliography{ISMIRtemplate}

%
%
%
%

\end{document}